\PassOptionsToPackage{unicode}{hyperref}
\PassOptionsToPackage{hyphens}{url}
\documentclass[12pt]{article}
\usepackage{amsmath,amssymb}
\usepackage{graphicx}
\usepackage{float}
\usepackage{iftex}
\ifPDFTeX
  \usepackage[T1]{fontenc}
  \usepackage[utf8]{inputenc}
  \usepackage{textcomp} % provide euro and other symbols
\else % if luatex or xetex
  \usepackage{unicode-math} % this also loads fontspec
  \defaultfontfeatures{Scale=MatchLowercase}
  \defaultfontfeatures[\rmfamily]{Ligatures=TeX,Scale=1}
\fi
\usepackage{lmodern}
\usepackage[margin=1in]{geometry}
\ifPDFTeX\else
\fi
\IfFileExists{upquote.sty}{\usepackage{upquote}}{}
\IfFileExists{microtype.sty}{% use microtype if available
  \usepackage[]{microtype}
  \UseMicrotypeSet[protrusion]{basicmath} % disable protrusion for tt fonts
}{}
\makeatletter
\@ifundefined{KOMAClassName}{% if non-KOMA class
  \IfFileExists{parskip.sty}{%
    \usepackage{parskip}
  }{% else
    \setlength{\parindent}{0pt}
    \setlength{\parskip}{6pt plus 2pt minus 1pt}}
}{% if KOMA class
  \KOMAoptions{parskip=half}}
\makeatother
\usepackage{xcolor}
\usepackage{booktabs,array}
\usepackage{caption}
\usepackage{fancyhdr}
\usepackage{calc} % for calculating minipage widths
\ifLuaTeX
  \usepackage{selnolig}  % disable illegal ligatures
\fi
\usepackage{bookmark}
\IfFileExists{xurl.sty}{\usepackage{xurl}}{} % add URL line breaks if available
\hypersetup{
  pdftitle={Is the Linear Threshold Good Enough? A Scale-Free Parameter and Adequacy Test for Curvature-Induced Threshold Displacement},
  pdfauthor={Subir Hait},
  hidelinks,
  pdfcreator={LaTeX}}

\begin{document}
\pagestyle{fancy}
\fancyhf{}
\fancyhead[R]{\thepage}
\setlength{\headheight}{15pt}
\renewcommand{\headrulewidth}{0pt}
\pagenumbering{arabic}
\thispagestyle{fancy}

\vspace*{0.16\textheight}
\begin{center}
{\LARGE\bfseries Is the Linear Threshold Good Enough?\par}
\vspace{0.45em}
{\Large\bfseries A Scale-Free Parameter and Adequacy Test for\par}
{\Large\bfseries Curvature-Induced Threshold Displacement\par}
\vspace{2.4em}
{\large Subir Hait\par}
\vspace{0.9em}
Department of Counseling, Educational Psychology, and Special Education (CEPSE)\\
College of Education, Michigan State University\\
620 Farm Lane, East Lansing, MI 48824, USA\\
\texttt{haitsubi@msu.edu}
\vfill
\end{center}

\newpage
\thispagestyle{fancy}
\begin{abstract}
Applied work routinely locates a threshold --- a break-even point, a
benchmark dose, a critical exposure --- by linearizing a smooth function
about a reference point and solving for the crossing. When the function
is curved the linear crossing is displaced, and the substantive
conclusion can change while every reported standard error remains valid.
We introduce the curvature-overstatement parameter
\(\Theta_{COT} = \log\left( \left| h_{2}^{*} \right|/\left| h_{1}^{*} \right| \right)\),
the log ratio of the second-order to the first-order threshold
displacement, and develop inference for it.

Our central structural result is an exact reduction. On the branch of
the quadratic root continuous with the linear solution,
\(\left| h_{2}^{*}/h_{1}^{*} \right| = 2/\left( 1 + \sqrt{1 - u} \right)\),
where \(u = 2qa/b^{2}\) is a single dimensionless index formed from the
local gap, slope, and curvature. Hence \(\Theta_{COT} = \phi(u)\)
depends on the three local parameters through one scalar, is invariant
to the units of both axes, and in the regular domain has range
\(\left( - \infty,\log 2 \right)\), with continuous boundary limit
\(\log 2\) at tangency. Thus linearization can understate the
displacement needed to reach the target by a factor approaching two, but
can overstate it without bound. We also show that the principal branch
is the globally nearest crossing, which removes a root-selection
ambiguity that informal treatments treat as a modeling choice, and that
the two regularity conditions usually stated separately---a real root
and a non-vanishing derivative at that root---are the same condition,
since \(F_{2}\prime\left( h_{2}^{*} \right) = \operatorname{sgn}(b)\sqrt{D}\).

We establish \(\sqrt{n}\)-normality with an explicit variance in the
regular case, and characterize two failures. Along local-to-tangency
sequences \(\sqrt{n}\left( 1 - u_{n} \right) \rightarrow \kappa\) the
estimator exists with probability \(\Phi(\kappa/\omega)\), converges at
rate \(n^{1/4}\), and has a non-normal limit; at \(\kappa = 0\) this
gives existence probability \(1/2\). As the slope vanishes the linear threshold becomes weakly identified, and the resulting ratio problem inherits the Gleser--Hwang pathology; short regular intervals cannot be uniformly trusted. A remainder theorem
bounds the distance between the second-order root and the true root by
\(M_{3}\left| h_{2}^{*} \right|^{3}/\left( 6\sqrt{D} \right)\), which
makes explicit both that \(\Theta_{COT}\) measures first- versus
second-order displacement rather than error against the truth, and that
tangency degrades the approximation itself and not merely its inference.

The practical contribution is an adequacy formulation that permits an
affirmative statement that the linear threshold is accurate within a
prespecified proportional tolerance, rather than treating failure to
detect curvature as evidence of adequacy. Under exact linearity, the
level-0.05 adequacy procedure affirms adequacy in 76.3\% of Monte Carlo
samples at n = 500 and essentially all samples at n = 2000, while it
does not affirm adequacy under the regular material-distortion
condition. The same study identifies near tangency and weak slope as
regimes in which regular inference should not be used. Together, the
effect-size scale, adequacy test, and diagnostics answer a practical
question that a sharp curvature test is not designed to answer: whether
a first-order threshold is accurate enough to report.

\end{abstract}


\newpage
\section{Introduction}\label{introduction}

Many applied procedures reduce to locating a point at which a smooth
function crosses a prespecified boundary. Break-even prices, benchmark
doses, effective concentrations, reliability limits, and critical values
of a sensitivity parameter share this structure: estimate a function,
fix a target, solve for the crossing.

One common way to solve for the crossing is to linearize. A first-order
expansion about a reference point has a closed-form root and propagates
uncertainty through the delta method with almost no effort. We are
careful about the scope of this claim. Modern benchmark-dose and
dose-response practice frequently inverts a fully specified nonlinear
model directly, using profile likelihood, model averaging, Bayesian, or
nonparametric methods, and does not linearize at all (Crump, 1984; Wheeler
and Bailer, 2007, 2008; Shao and Gift, 2014; Piegorsch, 2014; Lin et al.,
2015; Ritz et al., 2015; Jensen et al., 2019; Wheeler et al., 2022). The problem addressed here
arises in the narrower but still common situation in which a threshold
is \emph{reported or interpreted through} a first-order local
approximation --- marginal-effect extrapolations, tangent-based
break-even calculations, local sensitivity thresholds, screening
calculations, and the reading of a fitted slope as though it extended to
the target. In those settings the question of whether the linear answer
is good enough is live and, at present, has no formal answer.

Linearization misplaces the crossing whenever the function bends. If it
bends away from the boundary the linear solution reaches the target too
soon; if toward, too late. Either error changes the conclusion --- the
dose declared safe, the price at which a firm breaks even --- while
leaving every reported standard error intact. The failure is of
approximation, not precision.

The natural response is to test whether the quadratic coefficient is
zero. That answers a different question. Curvature is a property of the
function; threshold displacement is a property of the function
\emph{together with} the local slope and the distance to the target.
Strong curvature with a steep slope displaces the threshold negligibly.
Weak curvature with a shallow slope can displace it enormously. This
dependence is encoded by \(u = 2qa/b^{2}\).

A second limitation of the curvature test is directional. Failure to
reject \(q = 0\) is routinely read as licensing the linear threshold,
but a non-significant curvature coefficient is not evidence of adequacy
--- it is compatible with a wide range of threshold displacements,
especially at small \(n\). Establishing adequacy requires an
interval-null procedure, and that requires a parameter on which a
tolerance can be stated.

This paper supplies that parameter. Writing \(h_{1}^{*}\) and
\(h_{2}^{*}\) for the threshold displacements implied by the first- and
second-order expansions,

\[\Theta_{COT}\mspace{6mu} = \mspace{6mu}\log\left| \frac{h_{2}^{*}}{h_{1}^{*}} \right|,\tag{1}\]

a dimensionless quantity readable as a percentage:
\(\Theta_{COT} = - 0.323\) means the curvature-adjusted threshold sits
\(27.6\%\) nearer the reference point than linearization indicates.

\textbf{Sign convention.} Because ``overstatement'' flips meaning
according to whether one refers to distance or to speed of approach, we
fix one referent and use it throughout: \(\Theta_{COT}\) is a ratio of
\emph{distances}. Thus \(\Theta_{COT} > 0\) means the linear threshold
is too near the reference point, i.e.~linearization \textbf{understates
the displacement} required (equivalently, overstates the speed at which
the boundary is reached). \(\Theta_{COT} < 0\) means the linear
threshold is too far, i.e.~linearization \textbf{overstates the
displacement}. We retain the established acronym COT, in which
``overstatement'' refers to the speed reading, but every statement of
magnitude below is in distance units.

\textbf{Contribution.} We (i) define \(\Theta_{COT}\) and prove an exact
reduction to a one-dimensional curvature index, with invariance,
monotonicity, and an asymmetric range; (ii) resolve root selection by
continuity rather than by convention; (iii) derive regular asymptotics
and characterize the tangency and weak-slope regimes within one
framework; (iv) bound the gap between the second-order root and the true
root; and (v) convert the parameter into relevance and adequacy tests.

\section{Background and methodological context}\label{background-and-methodological-context}

\subsection{Threshold and inverse
estimation}\label{threshold-and-inverse-estimation}

Estimating an input value \(x^{*}\) defined implicitly by
\(m\left( x^{*} \right) = c\) spans inverse regression and calibration
(Osborne, 1991), benchmark-dose estimation (Crump, 1984;
Budtz-J{\o}rgensen, Keiding, and Grandjean, 2001; Wheeler and Bailer,
2007; Ritz and Streibig, 2005; Ritz et al., 2015; Jensen et al., 2019), inverse dose-response prediction (Demidenko et
al., 2013), effective-dose estimation, break-even analysis, and
level-crossing problems. This
literature is mature. It supplies Wald, likelihood-ratio,
profile-likelihood, bootstrap, Bayesian, and semiparametric inference
for the threshold itself, and modern practice often inverts the fitted
nonlinear model rather than a local linearization. We therefore make no
claim that thresholds lack inferential treatment. What this literature
does not supply is a scale-free comparison between the crossing implied
by a first-order approximation and the crossing implied by a
second-order one.

\subsection{Ratio inference and weak
denominators}\label{ratio-inference-and-weak-denominators}

The first-order threshold \(h_{1}^{*} = - a/b\) is a ratio estimator and
inherits the instability of ratio inference as \(b \rightarrow 0\).
Fieller (1954) provides the classical interval; Gleser and Hwang (1987)
show that when the denominator's limiting distribution has support
containing zero, no confidence set for the ratio with correct coverage
can have finite expected length; Hirschberg and Lye (2010) compare the
delta and Fieller geometries. We treat the weak-slope regime of Section
6.3 as an instance of this established pathology rather than as a new
discovery.

\subsection{Curvature testing and nonlinear Wald
inference}\label{curvature-testing-and-nonlinear-wald-inference}

Testing \(q = 0\) is a standard model-specification check. Because
\(\Theta_{COT}\) is a nonlinear transformation of \((a,b,q)\), the paper
must also engage the literature showing that Wald statistics are not
invariant under nonlinear reparameterization (Gregory and Veall, 1985;
Lafontaine and White, 1986; Phillips and Park, 1988). Accordingly, the
sharp COT null is treated as a specification check rather than as the
paper's primary inferential target; Section 8.2 compares it directly
with the ordinary curvature test.

\subsection{Nonlinearity diagnostics in nonlinear
regression}\label{nonlinearity-diagnostics-in-nonlinear-regression}

Classical nonlinear-regression diagnostics ask when curvature makes
linear inferential approximations unreliable. Beale (1960) proposed a
numerical measure of nonlinearity tied to the accuracy of approximate
confidence regions and to the effect of reparameterization. Bates and
Watts (1980) recast nonlinearity geometrically, separating intrinsic
curvature of the solution locus from parameter-effects curvature induced
by parameterization. Bates and Watts (1988) developed the geometric framework further, and
Ratkowsky (1983) developed related ideas as practical diagnostics for
assessing close-to-linear behavior in nonlinear regression. Box (1971)
linked nonlinear-estimation bias to these curvature ideas, while Seber
and Wild (1989) provide a broader treatment of curvature, reparameterization,
and nonlinear least-squares inference.

COT addresses a related but different object. Those diagnostics
characterize nonlinearity of a model-design-parameterization combination
and its consequences for parameter inference. COT instead takes a local
response-function expansion and a prespecified target and measures how
curvature displaces that level crossing relative to its first-order
location. Because threshold displacement also depends on the local gap
to the target and the local slope, inferential nonlinearity and
threshold distortion need not rank problems in the same way. The two
approaches are therefore complementary: classical curvature diagnostics
ask whether linear inferential geometry is trustworthy; COT asks whether
a linearized threshold is substantively adequate.

\subsection{Equivalence and relevance
testing}\label{equivalence-and-relevance-testing}

Deciding whether a parameter lies inside a prespecified tolerance is
standard equivalence testing, with two one-sided tests as the canonical
procedure (Schuirmann, 1987; Berger and Hsu, 1996; Wellek, 2010). We do not invent this logic.
Our use of it is to apply it to a newly defined estimand so that
adequacy of a first-order threshold becomes a rejectable hypothesis.

\subsection{Positioning of the contribution}\label{positioning-of-the-contribution}

Existing curvature tests determine whether a response function departs
from local linearity. Classical nonlinear-regression diagnostics
quantify intrinsic or parameter-effects nonlinearity and assess its
consequences for parameter inference. Existing threshold methods
estimate a crossing point and quantify its uncertainty. Ratio-inference
methods explain instability in linear threshold estimators. To our
knowledge, these literatures do not jointly provide (i) a scale-free
estimand for how much a first-order threshold moves once local curvature
is incorporated, (ii) a formal adequacy test asking whether a linear
threshold is accurate within a practical margin, (iii) an integrated
treatment of the regular, tangency, no-crossing, and weak-slope regimes
for that specific parameter, or (iv) guidance connecting the distortion
parameter to root selection and bootstrap validity.

This positioning is intentionally stated as a literature-based claim rather than as an
established nonexistence. Searches cannot establish absence,
particularly across older calibration, inverse-prediction,
numerical-analysis, and engineering literatures, and each ingredient
here has ancestors. The contribution is the synthesis and the estimand,
not the invention of threshold estimation, Taylor approximation, or
equivalence testing.

\section{Framework and definition}\label{framework-and-definition}

Let \(m(x;\psi)\) be three times continuously differentiable near a
reference point \(x_{0}\), indexed by
\(\psi \in \Psi \subseteq \mathbb{R}^{p}\), and fix a target \(c\).
Write \(h = x - x_{0}\) and define

\[a = m\left( x_{0};\psi \right) - c,\quad\quad b = \partial_{x}m\left( x_{0};\psi \right),\quad\quad q = \partial_{x}^{2}m\left( x_{0};\psi \right),\tag{2}\]

collected as \(\eta = (a,b,q)^{\top}\), so that

\[m\left( x_{0} + h;\psi \right) - c\mspace{6mu} = \mspace{6mu} a + bh + \frac{1}{2}qh^{2} + R(h),\quad\quad
R(h)=\frac{1}{2}\int_{0}^{h}(h-t)^{2}\,\partial_{x}^{3}m(x_{0}+t;\psi)\,dt.\tag{3}\]

Thus, whenever $|\partial_x^3m|\le M_3$ on the segment between $x_0$
and $x_0+h$, $|R(h)|\le M_3|h|^3/6$. The integral form is used below
because it also yields a valid derivative bound without differentiating
a Lagrange remainder point.

The factor \(\frac{1}{2}\) matters, because the convention propagates
into every later formula: with \(q\) the raw second derivative the
discriminant is \(b^{2} - 2qa\) and the root derivative involves
\(b + qh\), whereas writing \(a + bh + qh^{2}\) gives \(b^{2} - 4qa\)
and \(b + 2qh\). Both are internally consistent; mixing them is not.

Truncating (3) at first order gives the \textbf{linear threshold
displacement} \(h_{1}^{*} = - a/b\) for \(b \neq 0\). Truncating at
second order gives \(F_{2}(h): = a + bh + \frac{1}{2}qh^{2} = 0\) with
discriminant

\[D\mspace{6mu} = \mspace{6mu} b^{2} - 2qa.\tag{4}\]

\textbf{Definition 1 (principal branch).} The \emph{principal quadratic
threshold displacement} \(h_{2}^{*}\) is the root of \(F_{2}(h) = 0\)
converging to \(h_{1}^{*}\) as \(q \rightarrow 0\) with \((a,b)\) fixed;
for \(q = 0\) set \(h_{2}^{*} = h_{1}^{*}\).

\textbf{Definition 2.} For \(a \neq 0\), \(b \neq 0\), and \(D > 0\),
the curvature-overstatement parameter is
\(\Theta_{COT} = \log\left| h_{2}^{*}/h_{1}^{*} \right|\). At the
tangency boundary \(D = 0\) (equivalently \(u = 1\)), regular inference
does not apply; the continuous closure is
\({\bar{\Theta}}_{COT} = \log 2\).

\section{Structural theory}\label{structural-theory}

\subsection{Existence, the principal root, and a derivative
identity}\label{existence-the-principal-root-and-a-derivative-identity}

\textbf{Theorem 1.} Let \(a \neq 0\), \(b \neq 0\), and
\(u = 2qa/b^{2}\), so \(D = b^{2}(1 - u)\). If \(u < 1\) then the
principal root exists, is unique, and equals

\[h_{2}^{*}\mspace{6mu} = \mspace{6mu}\frac{- 2a}{\, b + \operatorname{sgn}(b)\sqrt{D}\,}.\tag{5}\]

It satisfies
\(\operatorname{sgn}\left( h_{2}^{*} \right) = \operatorname{sgn}\left( h_{1}^{*} \right)\), it is the
crossing of smallest absolute displacement among all real roots, and

\[F_{2}\prime\left( h_{2}^{*} \right)\mspace{6mu} = \mspace{6mu} b + q\, h_{2}^{*}\mspace{6mu} = \mspace{6mu} \operatorname{sgn}(b)\sqrt{D}.\tag{6}\]

\emph{Proof.} See Appendix~\ref{app:proofs}, Proof A.1.

Three consequences deserve emphasis. First, root selection is not a
modeling choice requiring pre-registration: continuity with the linear
solution selects the globally nearest crossing. Second, the concern that
the two approximations might point in opposite directions cannot arise
on the principal branch. Third, identity (6) shows that the two
regularity conditions usually stated separately --- existence of a real
root (\(D \geq 0\)) and non-vanishing of the derivative at that root
(\(b + qh_{2}^{*} \neq 0\)) --- are the same condition, \(D > 0\).
Equation (5) is also the numerically stable form; the naive quadratic
formula suffers catastrophic cancellation as \(q \rightarrow 0\).

\subsection{Exact dimensionless
representation}\label{exact-dimensionless-representation}

\textbf{Theorem 2.} Under the conditions of Theorem 1,

\[\frac{h_{2}^{*}}{h_{1}^{*}}\mspace{6mu} = \mspace{6mu}\frac{2}{1 + \sqrt{1 - u}}\mspace{6mu} > \mspace{6mu} 0,\quad\quad\text{hence}\quad\quad\Theta_{COT}\mspace{6mu} = \mspace{6mu}\phi(u)\mspace{6mu}: = \mspace{6mu}\log 2 - \log\left( 1 + \sqrt{1 - u} \right).\tag{7}\]

\emph{Proof.} See Appendix~\ref{app:proofs}, Proof A.2.

\textbf{Corollary 1 (dimensional reduction).} \(\Theta_{COT}\) depends
on \(\eta = (a,b,q)\) only through the scalar \(u\). Every inferential
problem for \(\Theta_{COT}\) is one-dimensional in \(u\).

\emph{Proof.} See Appendix~\ref{app:proofs}, Proof A.3.

\textbf{Corollary 2 (invariance).} \(u\), and hence \(\Theta_{COT}\), is
invariant under affine reparameterization
\(x \mapsto \alpha + \beta x\), \(\beta \neq 0\), and under rescaling
\(m - c \mapsto \lambda(m - c)\), \(\lambda \neq 0\).

\emph{Proof.} See Appendix~\ref{app:proofs}, Proof A.4.

Corollary 2 is what a distortion measure must satisfy: doses in
milligrams or grams, responses raw or standardized, must give the same
answer. None of \(q\), \(D\), or \(h_{2}^{*} - h_{1}^{*}\) has this
property, which is why the log ratio rather than a difference is the
right object.

\subsection{Monotonicity, range, and
sign}\label{monotonicity-range-and-sign}

\textbf{Theorem 3.} \(\phi\) is strictly increasing and continuously
differentiable on \(( - \infty,1)\) with

\[\phi\prime(u)\mspace{6mu} = \mspace{6mu}\frac{1}{2\sqrt{1 - u}\,\left( 1 + \sqrt{1 - u} \right)}\mspace{6mu} > \mspace{6mu} 0,\tag{8}\]

\(\phi(0) = 0\), \(\phi(u) \rightarrow - \infty\) as
\(u \rightarrow - \infty\), and \(\phi(u) \uparrow \log 2\) as
\(u \uparrow 1\). Hence \(\phi\) maps \(( - \infty,1)\) bijectively onto
\(\left( - \infty,\log 2 \right)\), with inverse
\(u = 1 - \left( 2e^{- \Theta} - 1 \right)^{2}\) for
\(\Theta < \log 2\). Its continuous extension to the boundary satisfies
\(\bar{\phi}(1) = \log 2\).

\emph{Proof.} See Appendix~\ref{app:proofs}, Proof A.5.

\textbf{Corollary 3 (asymmetric range).} In the regular domain,
\(0 < \left| h_{2}^{*}/h_{1}^{*} \right| < 2\). In the distance
convention of Section 1, a first-order approximation can understate the
displacement required to reach the target by a factor arbitrarily close
to two and can overstate it without bound. The factor two is the
supremum and is attained only in the continuous tangency closure.

\emph{Proof.} See Appendix~\ref{app:proofs}, Proof A.6.

The asymmetry is the most consequential structural fact about the
parameter and is invisible in Definition 2. Its upper limit is
approached only as \(u \uparrow 1\), precisely the tangency
configuration where regular inference fails and the second-order
approximation itself becomes fragile (Theorem 6 and Section 5).

\textbf{Corollary 4 (sign).}
\(\operatorname{sgn}\left( \Theta_{COT} \right) = \operatorname{sgn}(u) = \operatorname{sgn}(qa)\).

\emph{Proof.} See Appendix~\ref{app:proofs}, Proof A.7.

The direction of distortion depends only on whether curvature bends the
function toward or away from the target, not on the slope or the
direction of travel.

\subsection{What the sharp null is}\label{what-the-sharp-null-is}

\textbf{Proposition 1.} If \(a \neq 0\) and \(b \neq 0\), then
\(\Theta_{COT} = 0 \Leftrightarrow u = 0 \Leftrightarrow q = 0\).

\emph{Proof.} See Appendix~\ref{app:proofs}, Proof A.8.

We state this prominently because it disposes of the most natural
overclaim. Testing \(H_{0}:\Theta_{COT} = 0\) is testing \(q = 0\)
through a nonlinear reparameterization. By Wald non-invariance the two
statistics differ in finite samples, and Section 8.2 shows the
difference runs against the proposed statistic. The contribution lies in
magnitude, not in the sharp null.

\section{Approximation error against the true
threshold}\label{approximation-error-against-the-true-threshold}

\(\Theta_{COT}\) compares two approximations. It does not measure error
against the true crossing \(h^{*}\) solving
\(m\left( x_{0} + h^{*} \right) - c = 0\), and for a genuinely
non-quadratic \(m\) the two differ. The following makes the gap explicit
and bounds it.

\textbf{Theorem 4 (remainder bound).} Let \(D > 0\), let
\(J = \left\lbrack h_{2}^{*} - \Delta,\, h_{2}^{*} + \Delta \right\rbrack\),
and suppose \(\left| \partial_{x}^{3}m \right| \leq M_{3}\) on the
corresponding \(x\)-interval. Define

\[L(\Delta)\mspace{6mu} = \mspace{6mu}\sqrt{D}\mspace{6mu} - \mspace{6mu}|q|\Delta\mspace{6mu} - \mspace{6mu}\frac{1}{2}M_{3}(\left| h_{2}^{*} \right| + \Delta)^{2}.\tag{9}\]

If \(L(\Delta) > 0\) and
\(M_{3}\left| h_{2}^{*} \right|^{3}/6 \leq L(\Delta)\,\Delta\), then
\(F(h) = m\left( x_{0} + h \right) - c\) has exactly one zero \(h^{*}\)
in \(J\) and \(\left| h^{*} - h_{2}^{*} \right| \leq \Delta\).

\emph{Proof.} See Appendix~\ref{app:proofs}, Proof A.9.

\textbf{Corollary 5 (leading order).} When \(|q|\Delta\) and
\(\frac{1}{2}M_{3}\left( \left| h_{2}^{*} \right| + \Delta \right)^{2}\)
are small relative to \(\sqrt{D}\),

\[\left| h^{*} - h_{2}^{*} \right|\mspace{6mu} \lesssim \mspace{6mu}\frac{M_{3}\,\left| h_{2}^{*} \right|^{3}}{6\sqrt{D}}\mspace{6mu} = \mspace{6mu}\frac{M_{3}\,\left| h_{2}^{*} \right|^{3}}{6\,|b|\sqrt{1 - u}},\tag{10}\]

and consequently
\(|\log\left| h^{*}/h_{1}^{*} \right| - \Theta_{COT}| \leq - \log(1 - \rho)\)
with \(\rho = M_{3}\left| h_{2}^{*} \right|^{2}/\{ 6|b|\sqrt{1 - u}\}\),
provided \(\rho < 1\).

\emph{Proof.} See Appendix~\ref{app:proofs}, Proof A.10.

Three implications. First, \(\Theta_{COT}\) should be read as
\emph{first- versus second-order displacement}, and the paper claims
nothing more; (10) is the price of that restriction and the route to a
stronger claim when \(M_{3}\) is available. Second, the bound degrades
as \(u \uparrow 1\): \textbf{tangency corrupts the second-order
approximation itself, not merely its sampling behavior}, which unifies
this section with Theorem 6. Third, for a quadratic model \(M_{3} = 0\) and
\(h_{2}^{*} = h^{*}\) exactly --- which is why the Monte Carlo design of
Section 8 isolates inferential behavior from approximation error. In
practice we recommend reporting \(\widehat{\rho}\) alongside
\(\widehat{\Theta}\) whenever a bound on the third derivative is
available.

\section{Sampling theory}\label{sampling-theory}

Let \(\widehat{\eta}\) satisfy
\(\sqrt{n}\left( \widehat{\eta} - \eta_{0} \right) \Rightarrow N(0,\Sigma)\)
with \(\Sigma\) positive definite and \(\widehat{\Sigma}\) consistent
--- true for maximum likelihood, nonlinear least squares, GEE, and
sandwich estimation --- and write
\(\widehat{u} = 2\widehat{q}\widehat{a}/{\widehat{b}}^{2}\),
\(\widehat{\Theta} = \phi\left( \widehat{u} \right)\).

\subsection{Regular case}\label{regular-case}

\textbf{Theorem 5.} If \(a_{0} \neq 0\), \(b_{0} \neq 0\), and
\(u_{0} < 1\), then
\(\sqrt{n}\left( \widehat{\Theta} - \Theta_{0} \right) \Rightarrow N\left( 0,\sigma^{2} \right)\)
with

\[\sigma^{2} = \phi\prime\left( u_{0} \right)^{2}\,\nabla u\left( \eta_{0} \right)^{\top}\Sigma\,\nabla u\left( \eta_{0} \right),\quad\quad\nabla u(\eta) = \left( \frac{2q}{b^{2}},\, - \frac{4qa}{b^{3}},\,\frac{2a}{b^{2}} \right)^{\top},\tag{11}\]

and \({\widehat{\sigma}}^{2}\) obtained by substitution is consistent.

\emph{Proof.} See Appendix~\ref{app:proofs}, Proof A.11.

When \(a_{0},q_{0} \neq 0\),
\(\nabla u = u \cdot (1/a,\, - 2/b,\, 1/q)^{\top}\), so
\[
\sigma^{2}=\{u_{0}\phi'(u_{0})\}^{2}\,
\operatorname{avar}\!\left(\log|\widehat q|+\log|\widehat a|-2\log|\widehat b|\right).
\]
Thus
the variance of \(\widehat{\Theta}\) is the variance of
\(\log\left| \widehat{u} \right|\) scaled by the squared elasticity of
\(\phi\). Precision therefore requires precision in all three local
parameters on the log scale; a well-estimated curvature does not rescue
a poorly estimated slope.

\subsection{Tangency and local-to-tangency
sequences}\label{tangency-and-local-to-tangency-sequences}

This is a boundary-type nonregular problem. Chernoff (1954), Self and
Liang (1987), and Andrews (1999) provide general background on boundary
asymptotics, although the $n^{1/4}$ transformation below is specific to
the present level-crossing parameter.

\textbf{Theorem 6.} Let $\eta_n=(a_n,b_n,q_n)^\top\to\eta_0$
with $a_0\neq0$, $b_0\neq0$, and $u(\eta_0)=1$. Suppose the triangular
array satisfies
$\sqrt n(\widehat\eta_n-\eta_n)\Rightarrow N(0,\Sigma)$ and
$\widehat\Sigma\to_p\Sigma$, and let
$u_n=u(\eta_n)$ obey $\sqrt n(1-u_n)\to\kappa\ge0$. Put
$\omega^2=\nabla u(\eta_0)^\top\Sigma\nabla u(\eta_0)>0$. Then:

\begin{enumerate}
\def\labelenumi{(\roman{enumi})}
\item
  $\Pr(\widehat\Theta\ \text{defined})=\Pr(\widehat u<1)\to
  \Phi(\kappa/\omega)$;
\item
  conditionally on $\widehat u<1$,
  $n^{1/4}(\log2-\widehat\Theta)\Rightarrow\sqrt V$, where
  $V=\kappa-\omega Z$ and $Z\sim N(0,1)$ is conditioned on $V>0$;
\item
  if $\widehat{se}(\widehat\Theta)=\widehat\omega\,
  \phi'(\widehat u)/\sqrt n$ denotes the regular delta standard error,
  then, conditionally on $\widehat u<1$,
  \[
  n^{1/4}\widehat{se}(\widehat\Theta)\Rightarrow
  \frac{\omega}{2\sqrt V},\qquad
  \frac{\widehat\Theta-\Theta_n}{\widehat{se}(\widehat\Theta)}
  \Rightarrow \frac{2\sqrt V(\sqrt\kappa-\sqrt V)}{\omega},
  \]
  a non-Gaussian limit. Thus regular Gaussian delta calibration is not
  valid along the local-to-tangency sequence.
\end{enumerate}

At $\kappa=0$ this gives existence probability $1/2$, rate $n^{1/4}$,
and limit $\sqrt{\omega|Z|}$ with $|Z|$ half-normal (conditional on the
defined branch).

\emph{Proof.} See Appendix~\ref{app:proofs}, Proof A.12.

Because
\[
\mathbb{E}|Z|^{1/2}=\frac{2^{1/4}\Gamma(3/4)}{\sqrt{\pi}}\approx0.8222,
\]
the \(\kappa=0\) case predicts
\[
n^{1/4}\mathbb{E}\!\left[\log2-\widehat{\Theta}\mid\widehat u<1\right]
\longrightarrow0.8222\sqrt{\omega}.
\]
Table 2 closely agrees with part (i) across
\(\kappa \in \{0,1,3,6\}\), within Monte Carlo error; Table 3 illustrates
the predicted $n^{1/4}$ rate and the failure of regular Gaussian calibration
at exact tangency.

The practical implication is diagnostic, not corrective: report the
studentized discriminant
\(\sqrt{n}\left( 1 - \widehat{u} \right)/\widehat{\omega}\), which
estimates \(\kappa/\omega\), and when it is small do not present a
delta-method interval as if Theorem 5 applied.

\subsection{Weak slope}\label{weak-slope}

\textbf{Theorem 7.} Fix \(a \neq 0\), \(q \neq 0\) and let
\(b \rightarrow 0\). (i) If \(qa > 0\) then \(D \rightarrow - 2qa < 0\)
and for small \(|b|\) no real crossing exists. (ii) If \(qa < 0\) then
\(h_{1}^{*} \rightarrow \pm \infty\) while \(h_{2}^{*}\) remains finite,
and

\[\Theta_{COT} = \log 2 + \log|b| - \frac{1}{2}\log\left( 2|qa| \right) + o(1)\mspace{6mu} \rightarrow \mspace{6mu} - \infty.\tag{12}\]

\emph{Proof.} See Appendix~\ref{app:proofs}, Proof A.13.

The weak-slope problem is not that both thresholds destabilize --- the
quadratic threshold is well behaved --- but that the \emph{linear}
threshold diverges, so the defining ratio has an exploding denominator.
\(\Theta_{COT}\) is not locally identified at \(b = 0\).

\textbf{Remark 1.} Since \(h_{1}^{*} = - a/b\) is a ratio whose
denominator can have a limiting distribution with support containing
zero, the first-order threshold falls directly within the
Gleser--Hwang (1987) unbounded-confidence-set pathology. Because
\(\Theta_{COT}\) is defined through that threshold and diverges as
\(b\to0\), reparameterization does not remove the weak-denominator
problem. We therefore do not interpret short regular intervals near
\(b=0\) as evidence of precision; the slope diagnostic and the
invalid-root/bootstrap-failure diagnostics must accompany the estimate.

\section{Relevance and adequacy
inference}\label{relevance-and-adequacy-inference}

The sharp null is available but, by Proposition 1 and Section 8.2,
should be tested on \(q\). The two useful COT formulations fix a margin
\(\delta > 0\) on the log-ratio scale. Because
\(e^{\Theta_{COT}} = \left| h_{2}^{*}/h_{1}^{*} \right|\), choosing
\(\delta = \log(1.10) \approx 0.0953\) treats threshold ratios between
\(1/1.10\) and \(1.10\) as practically equivalent to one. This margin is
symmetric on the log-ratio scale, not on the \(u\) scale.

\textbf{Material distortion (relevance).} The scientific null is
\(H_{0}: - \delta \leq \Theta_{COT} \leq \delta\) against
\(H_{A}:\Theta_{COT} < - \delta\ \text{or}\ \Theta_{COT} > \delta\). The
positive and negative directional tests reject when
\(\widehat{\Theta} - z_{1 - \alpha}se > \delta\) and
\(\widehat{\Theta} + z_{1 - \alpha}se < - \delta\), respectively. Each
directional claim is a level-\(\alpha\) test. If direction was not
prespecified and a single omnibus level-\(\alpha\) relevance decision is
required, use \(\alpha/2\) in each tail, equivalently replacing
\(z_{1 - \alpha}\) by \(z_{1 - \alpha/2}\). Rejection licenses the
corresponding claim that linearization materially misplaces the
threshold.

\textbf{First-order adequacy (equivalence).} Test
\(H_{0}:\Theta_{COT} \leq - \delta\ \text{or}\ \Theta_{COT} \geq \delta\)
against \(H_{A}: - \delta < \Theta_{COT} < \delta\). The
level-\(\alpha\) TOST rejects when both
\(\left( \widehat{\Theta} + \delta \right)/se > z_{1 - \alpha}\) and
\(\left( \widehat{\Theta} - \delta \right)/se < - z_{1 - \alpha}\)
(Schuirmann, 1987). Rejection licenses the affirmative claim that the
linear threshold is usable within tolerance; failure to reject this
inadequacy null never licenses that claim. We regard this as the most
useful procedure in the paper.

\textbf{Proposition 2 (admissible margins).} For \(\delta \geq \log 2\),
no regular principal-branch value can satisfy \(\Theta_{COT} > \delta\).
Consequently the positive-direction relevance alternative is empty, and
the upper inadequacy component \(\Theta_{COT} \geq \delta\) is excluded
by the regular parameter space.

\emph{Proof.} See Appendix~\ref{app:proofs}, Proof A.14.

To retain two-sided scientific meaning, margins should therefore satisfy
\(0 < \delta < \log 2\). Values such as \(\log(1.05)\) through
\(\log(1.25)\) provide an interpretable illustrative range. By Theorem 3
the positive and negative margins map monotonically to the \(u\) scale
as \(u_{\delta}^{+} = 1 - \left( 2e^{- \delta} - 1 \right)^{2}\) and
\(u_{\delta}^{-} = 1 - \left( 2e^{\delta} - 1 \right)^{2}\); for
\(\delta = \log(1.10)\) these are \(0.331\) and \(- 0.440\). The mapping
is asymmetric, so a symmetric margin on the log-ratio scale imposes
different tolerances on the two sides of \(u = 0\).

\textbf{Algorithm 1 (procedure).} Given \(\psi\), \(Var(\psi)\),
\(x_{0}\), \(c\), \(\delta\), and \(\alpha\): (1) form \(\eta\) and
\(\Sigma = JVar(\psi)J^{\top}\) with
\(J = \partial\eta/\partial\psi^{\top}\); (2) report
\(t_{b} = \left| \widehat{b} \right|/se\left( \widehat{b} \right)\) as a
weak-slope diagnostic, but do not use any fixed cutoff as the sole
validity gate; (3) compute \(\widehat{u}\), \(\widehat{\omega}\), and
the studentized distance to tangency
\(T_{D} = \sqrt{n}\left( 1 - \widehat{u} \right)/\widehat{\omega}\); (4)
if \(\widehat{u} > 1\), report no real quadratic crossing, while
\(\widehat{u} = 1\) is the tangency boundary; (5) when bootstrapping,
report the invalid-root fraction
\(p_{fail}^{*} = B^{- 1}\sum_{b = 1}^{B}I\left( {\widehat{u}}_{b}^{*} \geq 1 \right)\).
If \(T_{D}\) is small or \(p_{fail}^{*}\) is non-negligible, flag
nonregularity and do not base relevance or adequacy claims on the delta
interval; a percentile-bootstrap interval may be reported only as a
finite-sample sensitivity analysis together with \(p_{fail}^{*}\); (6)
in the regular region report
\(\widehat{\Theta} = \phi\left( \widehat{u} \right)\),
\(se = \phi\prime\left( \widehat{u} \right)\widehat{\omega}/\sqrt{n}\),
the interval and its exponentiated version, the relevance and adequacy
decisions, \(x_{1}^{*}\), \(x_{2}^{*}\), and, where available, the
remainder ratio \(\rho\) of Corollary 5.

The percentile bootstrap is therefore a finite-sample sensitivity tool,
not a proved boundary correction. Always report the fraction of
resamples with \({\widehat{u}}^{*} \geq 1\). Table 4 shows improved
coverage in the simulated near-tangency condition, but
ordinary-bootstrap validity at exact tangency is not established here.
Do not use the basic (reflected) interval: because the regular parameter
space has supremum \(\log 2\), reflection can place the upper endpoint
outside even its continuous closure. No universal cutoff for \(T_{D}\)
or \(p_{fail}^{*}\) is established here, so these diagnostics should be
reported quantitatively rather than converted into a new binary test.

\section{Monte Carlo study}\label{monte-carlo-study}

\subsection{Design}\label{design}

The data-generating process is
\(y_{i} = a + bx_{i} + \frac{1}{2}qx_{i}^{2} + \varepsilon_{i}\) with
\(x_{i} \sim Uniform( - 1.5,1.5)\),
\(\varepsilon_{i} \sim N\left( 0,{0.5}^{2} \right)\), \(x_{0} = 0\),
\(c = 0\), estimated by ordinary least squares with
\(\widehat{\Sigma} = J\widehat{Var}\left( \widehat{\beta} \right)J^{\top}\),
\(J = diag(1,1,2)\). The expansion is exact, so by Corollary 5 every
departure from nominal behavior is attributable to the nonlinearity of
\(\Theta_{COT}\) in \(\eta\) rather than to approximation error. Four
conditions target the theorems directly; \(n \in \{ 100,500,2000\}\);
5000 replications per cell; \(\delta = \log(1.10)\), \(\alpha = .05\).

\clearpage
\begin{table}[!ht]
\caption{Four Monte Carlo conditions targeting Theorems 5, 6, and 7.}
\label{tab:mainmc}
\centering
\normalsize
\textit{Note.} median $t(b)$ is the median studentized slope; undef is the percentage with $\widehat u\ge1$; relev is the percentage for which either level-.05 directional relevance test of Section 7 rejects; equiv is the level-.05 adequacy test. Coverage and tests condition on definedness.\\[0.8em]
\textit{Panel A. Estimation behavior}\\[0.35em]
\begin{tabular}{lrrrrrr}
\toprule
Condition & $\theta_0$ & $n$ & median $t(b)$ & undef \% & bias & SD \\
\midrule
C1: null ($q=0$) & 0.0000 & 100  & 17.1 & 0.0  & 0.0066  & 0.0799 \\
                  & 0.0000 & 500  & 38.6 & 0.0  & 0.0007  & 0.0335 \\
                  & 0.0000 & 2000 & 77.4 & 0.0  & 0.0001  & 0.0170 \\
C2: regular ($u_0=.60$) & 0.2031 & 100  & 17.2 & 6.4  & -0.0013 & 0.1282 \\
                         & 0.2031 & 500  & 38.6 & 0.0  & 0.0072  & 0.0646 \\
                         & 0.2031 & 2000 & 77.4 & 0.0  & 0.0017  & 0.0306 \\
C3: near tangency ($u_0=.95$) & 0.4913 & 100  & 17.5 & 40.9 & -0.1663 & 0.1335 \\
                              & 0.4913 & 500  & 39.0 & 34.3 & -0.0731 & 0.0963 \\
                              & 0.4913 & 2000 & 77.6 & 19.7 & -0.0210 & 0.0752 \\
C4: weak slope ($b=.05$) & 0.2031 & 100  & 0.9 & 49.7 & -1.8653 & 1.0225 \\
                          & 0.2031 & 500  & 2.0 & 49.7 & -1.5594 & 0.8304 \\
                          & 0.2031 & 2000 & 3.9 & 49.7 & -1.1384 & 0.4534 \\
\bottomrule
\end{tabular}

\vspace{1em}
\textit{Panel B. Standard errors, coverage, and decisions}\\[0.35em]
\begin{tabular}{lrrrrr}
\toprule
Condition & $n$ & mean SE & cov 95 & relev & equiv \\
\midrule
C1: null ($q=0$) & 100  & 0.0802 & 0.963 & 0.002 & 0.003 \\
                  & 500  & 0.0337 & 0.952 & 0.000 & 0.763 \\
                  & 2000 & 0.0167 & 0.947 & 0.000 & 1.000 \\
C2: regular ($u_0=.60$) & 100  & 0.1787 & 0.927 & 0.002 & 0.000 \\
                         & 500  & 0.0642 & 0.960 & 0.594 & 0.000 \\
                         & 2000 & 0.0303 & 0.952 & 0.994 & 0.000 \\
C3: near tangency ($u_0=.95$) & 100  & 0.3948 & 0.771 & 0.002 & 0.000 \\
                              & 500  & 0.2261 & 0.866 & 0.811 & 0.000 \\
                              & 2000 & 0.1308 & 0.908 & 0.948 & 0.000 \\
C4: weak slope ($b=.05$) & 100  & 5.0387 & 0.817 & 0.169 & 0.000 \\
                          & 500  & 2.5972 & 0.591 & 0.391 & 0.000 \\
                          & 2000 & 1.4736 & 0.529 & 0.430 & 0.000 \\
\bottomrule
\end{tabular}
\end{table}
\clearpage

C1 confirms Theorem 5: negligible bias, analytic standard error matching
the Monte Carlo standard deviation, coverage \(0.947\)-\(0.963\), and
the adequacy test correctly establishing adequacy in \(76.3\%\) of samples
at \(n = 500\) and essentially all samples at \(n = 2000\)---the
affirmative conclusion no curvature test can deliver. C2 is the regular
case with material distortion; the directional relevance decision rises
from \(0.002\) to \(0.994\) while the adequacy test correctly never
fires. C3 shows the tangency degradation of Theorem 6 receding only as
\(n\) separates \(\widehat{u}\) from the boundary, with the mean
analytic standard error exceeding the Monte Carlo standard deviation by
about a factor of three at \(n = 100\).

\begin{figure}[H]
\centering
\includegraphics[width=0.88\textwidth]{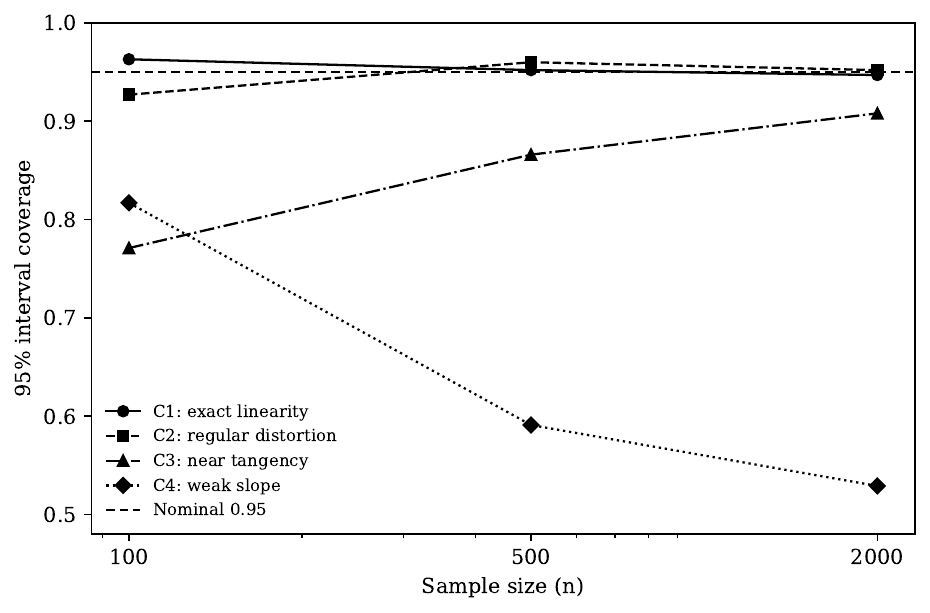}
\caption{Monte Carlo coverage of the nominal 95\% interval across the four conditions in Table~1. The horizontal reference line marks 0.95.}
\label{fig:coverage}
\end{figure}

C4 is the condition in which the method performs poorly over the sample
sizes studied, and we report it in full. The true distortion is
\(+ 0.203\); estimates remain severely negatively biased, coverage falls
to about \(0.53\), and the undefined fraction is \(49.7\%\) at each
simulated \(n\). Because sensitivity of \(u\) to the slope scales as
\(b^{-3}\), the weak slope creates extreme finite-sample instability here.
Since \(b = 0.05 \neq 0\), Theorem 5 still implies eventual regular
asymptotics; the problem is that convergence is too slow over the sample
sizes studied. Worse, the directional relevance decision rejects in
\(17\)-\(43\%\) of defined samples, and it does so in the wrong
direction: it declares material distortion of the opposite sign. This is
the Gleser--Hwang pathology of Remark 1 made concrete. The \(t_{b} = 3\)
guard trips at \(n = 100\) and \(n = 500\) but passes at \(n = 2000\),
where coverage is still \(0.529\); therefore no fixed \(t_{b}\) cutoff
is used as a sufficient regularity gate in Algorithm 1. The persistent
near-one-half invalid-root fraction is the more revealing signal,
motivating joint reporting of slope, discriminant, and bootstrap-failure
diagnostics. Figure~\ref{fig:coverage} summarizes the contrasting coverage
patterns across the four conditions.

\subsection{The sharp null}\label{the-sharp-null}

For completeness we compared the Wald test of \(H_{0}:\Theta_{COT} = 0\)
with the ordinary test of \(q = 0\) across the regular conditions. In the
comparisons examined, the curvature test had higher rejection rates: at
\(u_{0} = 0.6\), \(n = 100\), rates were \(0.058\) and \(0.484\) for
the COT and curvature tests, respectively; at \(u_{0} = 0.3\),
\(n = 250\), they were \(0.253\) and \(0.355\). This is Wald
non-invariance under a strongly nonlinear reparameterization and the
empirical counterpart of Proposition 1. \textbf{The sharp null should be
tested on} \(q\)\textbf{.} The COT machinery is for magnitude.

\clearpage
\subsection{Tangency predictions}\label{tangency-predictions}
\begin{table}[!ht]
\caption{Local-to-tangency existence probabilities.}
\label{tab:tangencyprob}
\centering
\normalsize
\textit{Note.} $u_n=1-\kappa/\sqrt n$, 6000 replications, $\omega=2.730$. Theorem 6(i) predicts $\Pr(\widehat u<1)\to\Phi(\kappa/\omega)$.\\[0.8em]
\begin{tabular}{rrrr}
\toprule
$\kappa$ & $n=1000$ & $n=4000$ & $\Phi(\kappa/\omega)$ \\
\midrule
0 & 0.503 & 0.505 & 0.500 \\
1 & 0.639 & 0.642 & 0.643 \\
3 & 0.859 & 0.866 & 0.864 \\
6 & 0.986 & 0.988 & 0.986 \\
\bottomrule
\end{tabular}
\end{table}
\clearpage

\begin{figure}[H]
\centering
\includegraphics[width=0.88\textwidth]{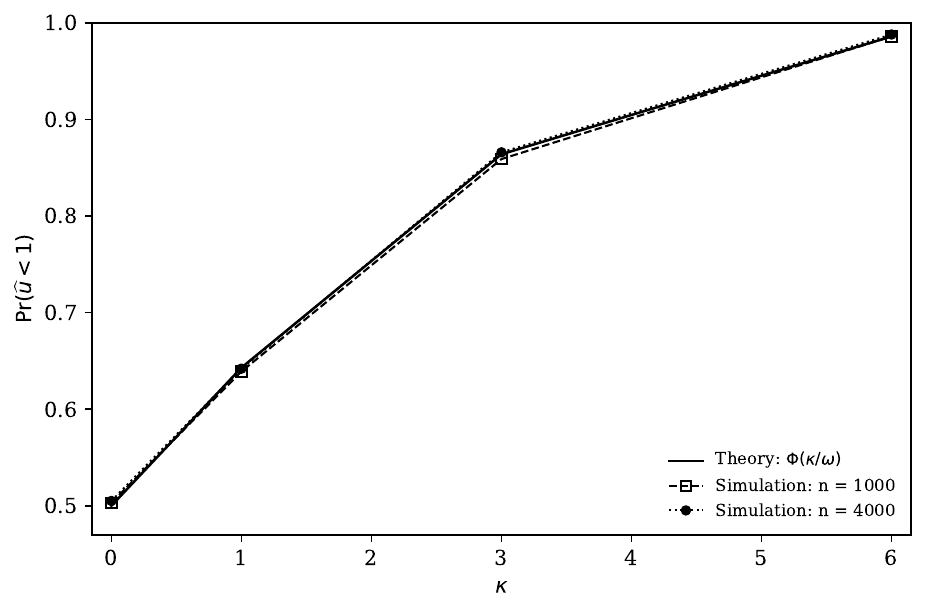}
\caption{Local-to-tangency existence probabilities from Table~2. The two simulated sequences nearly coincide with the theoretical limit $\Phi(\kappa/\omega)$.}
\label{fig:tangency}
\end{figure}

Figure~\ref{fig:tangency} visualizes the agreement between the finite-sample
simulation and the local-to-tangency probability limit.

\clearpage
\begin{table}[!ht]
\caption{Exact-tangency scaling and interval coverage.}
\label{tab:exacttangency}
\centering
\normalsize
\textit{Note.} $\kappa=0$; 3000 replications. Theory predicts a stable $n^{1/4}$-scaled mean, a diverging $n^{1/2}$-scaled mean, and non-Gaussian regular calibration.\\[0.8em]
\begin{tabular}{rrrrrr}
\toprule
$n$ & undef \% & mean $(\log2-\widehat\Theta)$ & $n^{1/4}\!\times$mean & $n^{1/2}\!\times$mean & cov 95 \\
\midrule
250  & 49.6 & 0.2888 & 1.148 & 4.566  & 0.596 \\
1000 & 49.6 & 0.2114 & 1.189 & 6.684  & 0.621 \\
4000 & 49.3 & 0.1545 & 1.229 & 9.774  & 0.638 \\
8000 & 48.6 & 0.1340 & 1.267 & 11.985 & 0.618 \\
\bottomrule
\end{tabular}
\end{table}
\clearpage

Across the eight simulated cells, the observed existence probabilities
closely track \(\Phi(\kappa/\omega)\), with deviations consistent with
Monte Carlo error. In Table 3 the \(n^{1/4}\)-scaled mean is stable near
\(1.2\) while the \(n^{1/2}\)-scaled mean grows by a factor of \(2.6\);
the predicted limit \(0.8222\sqrt{\omega} = 1.358\) is approached from
below at the slow rate the \(n^{1/4}\) scaling implies.

\clearpage
\subsection{Supplementary: bootstrap}\label{supplementary-bootstrap}
\begin{table}[!ht]
\caption{Bootstrap and delta-method interval coverage.}
\label{tab:bootstrap}
\centering
\normalsize
\textit{Note.} 95\% interval coverage, 600 replications, $B=399$ pairs resamples.\\[0.8em]
\begin{tabular}{rrrrrr}
\toprule
$u_0$ & $n$ & delta & percentile & basic & boot fail \% \\
\midrule
0.0 & 250 & 0.948 & 0.935 & 0.942 & 0.0 \\
0.6 & 500 & 0.945 & 0.940 & 0.915 & 0.9 \\
0.9 & 500 & 0.898 & 0.948 & 0.734 & 15.2 \\
\bottomrule
\end{tabular}
\end{table}
\clearpage

For general bootstrap and interval background see Efron and Tibshirani
(1993), Hall (1992), Davison and Hinkley (1997), and DiCiccio and Efron
(1996). Boundary problems can invalidate ordinary bootstrap
approximations (Andrews, 2000). Away
from tangency the methods are similar and the delta method is preferable
on computational cost. In the simulated near-tangency condition, the
percentile bootstrap improves coverage substantially and respects the
compression of the parameter space near \(\log 2\) better than a
symmetric Wald interval. This is empirical evidence, not a proof of
ordinary-bootstrap validity at the boundary. The basic interval performs
poorly for the reason Theorem 3 predicts: reflection about
\(\widehat{\Theta}\) can push the upper limit past \(\log 2\), outside
the continuous parameter-space closure.

\section{Numerical illustration}\label{numerical-illustration}

Consider a dose-response model
\(m(d) = \{ 1 + \exp\left\lbrack - \left( \beta_{0} + \beta_{1}d \right) \right\rbrack\}^{- 1}\)
for the probability of an adverse response, with a benchmark target of
\(c = 0.10\), reference dose \(d_{0} = 0\), and
\(\left( \beta_{0},\beta_{1} \right) = ( - 3,\, 0.8)\). Then
\(a = - 0.0526\), \(b = 0.0361\), \(q = 0.0262\), so \(u = - 2.107\) and

\[{\widehat{\Theta}}_{COT}\mspace{6mu} = \mspace{6mu} - 0.323,\quad\quad e^{\widehat{\Theta}} = 0.724\mspace{6mu}\mspace{6mu}( - 27.6\%).\]

The linear benchmark dose is \(h_{1}^{*} = 1.455\); the
curvature-adjusted value is \(h_{2}^{*} = 1.053\). Because this model
has a closed-form root, we can also report the truth: \(h^{*} = 1.004\).
The first-order calculation overstates the benchmark dose by \(45.0\%\)
relative to the true root; the second-order calculation is within
\(4.9\%\). Since the signed value above is negative, the tolerance
comparison uses its magnitude:
\(\left| \Theta_{COT} \right| = 0.323 > \delta = \log(1.10)\), so the
point-estimate displacement exceeds the prespecified practical tolerance
in magnitude. This numerical illustration supplies no sampling
covariance matrix, however, so no p-value or rejection claim is made.
Substantively, a regulator using the tangent-based dose in this example
would set a limit nearly half again too permissive. The residual
\(4.9\%\) gap between \(h_{2}^{*}\) and \(h^{*}\) is exactly the
quantity bounded by Theorem 4 and illustrates why \(\Theta_{COT}\) must
be read as first- versus second-order displacement rather than as total
approximation error.

\section{Discussion}\label{discussion}

The curvature-overstatement parameter is far simpler than its definition
suggests. It is a strictly increasing function of one dimensionless
index, invariant to the units of both axes, with regular range
\(\left( - \infty,\log 2 \right)\) and supremum \(\log 2\) attained only
in the continuous tangency closure, signed by the product of curvature
and the initial gap, with root selection fixed by continuity rather than
convention and with its two regularity conditions collapsing into one.
These facts do most of the work: they reduce the delta method to one
dimension, explain why reflected bootstrap intervals fail, locate the
exact configuration at which the asymptotics break, and bound the
admissible equivalence margin.

We have been deliberate about limits. The sharp null is not new, and its
Wald test had lower power than the ordinary curvature test in the comparisons
reported. Weak identification at \(b = 0\) can place the problem in the
Gleser--Hwang regime when the limiting denominator distribution has support
containing zero; reparameterization does not restore identification in that
regime. Condition C4 shows the resulting instability producing confidently
wrong-signed conclusions. The parameter measures first- versus second-order
displacement, not error against the truth, with Theorem 4 as the bridge. And the domain is
restricted: COT applies where a threshold is reported or interpreted
through a first-order approximation, or where adequacy of such an
approximation must be assessed. It does not supersede direct inversion
of a fitted nonlinear model, which is preferable whenever available
(Ritz and Streibig, 2005; Ritz et al., 2015; Jensen et al., 2019; Wheeler
et al., 2022).

What remains is a well-behaved effect-size scale for a question applied
work asks constantly and answers badly: whether a linear threshold may
be reported as it stands. On that question the adequacy formulation
delivers an affirmative answer with controlled error, which no test of
\(q = 0\) can provide.

Extensions. A third-order version would replace the quadratic root by a
cubic and sharpen Theorem 4, at the cost of the closed form in Theorem
2. The \(n^{1/4}\) rate at tangency motivates studying subsampling or an
\(m\)-out-of-\(n\) bootstrap; ordinary-bootstrap validity at the
boundary is left open here. The most substantial direction is geometric:
with a vector argument the target becomes a level set, the scalar \(q\)
becomes a Hessian, and the threshold becomes a distance to a quadric
surface. We flag one finding that argues for treating this as separate
work rather than a corollary. Along any fixed ray \(h = tv\) the entire
scalar theory transfers verbatim with \(b_{v} = g^{\top}v\) and
\(q_{v} = v^{\top}Hv\), so the directional field
\(v \mapsto \phi\left( 2av^{\top}Hv/\left( g^{\top}v \right)^{2} \right)\)
inherits Theorems 1-3 including the \(\log 2\) supremum. But the
nearest-distance generalization
\(\log\left( d_{2}^{*}/d_{1}^{*} \right)\) is a ratio of minima rather
than a minimum of ratios, and the \(\log 2\) bound survives only when
the gradient ray crosses the quadric; we constructed a counterexample
with \(d_{2}^{*}/d_{1}^{*} = 2.275\) when it does not. The multivariate
object is also metric-dependent, since Euclidean distance is not
invariant to coordinate rescaling, so Corollary 2 fails unless a metric
such as \(\Sigma_{x}^{- 1}\) is adopted. A geometric extension is
therefore a genuine research problem, not an immediate corollary, and
importing it into this paper would import an error.

\clearpage
\appendix
\section{Mathematical proofs}\label{app:proofs}

This appendix gives complete proofs of the structural and sampling results used in the main text. Throughout, $u=2qa/b^2$ and $D=b^2(1-u)$ whenever $b\neq0$.

\subsection*{Proof A.1: Theorem 1}
For $q\neq0$, the two quadratic roots are $(-b\pm\sqrt D)/q$. Define
\[
h_s=\frac{-b+\operatorname{sgn}(b)\sqrt D}{q}.
\]
As $q\to0$ with $(a,b)$ fixed,
$\sqrt D=|b|\{1-qa/b^2+O(q^2)\}$, so the numerator of $h_s$ equals
$-qa/b+O(q^2)$ and $h_s\to-a/b=h_1^*$. The other root diverges as
$-2b/q$. Rationalizing $h_s$ gives
\[
h_s=\frac{-2a}{b+\operatorname{sgn}(b)\sqrt D},
\]
which is (5) and extends continuously to $q=0$.
Because $\sqrt D=|b|\sqrt{1-u}$, the denominator in (5) is
$b\{1+\sqrt{1-u}\}$ and therefore has the sign of $b$. Hence
$\operatorname{sgn}(h_2^*)=\operatorname{sgn}(-a/b)=\operatorname{sgn}(h_1^*)$.
For the other real root $h_o$,
\[
|h_2^*|=\frac{2|a|}{|b|+\sqrt D},\qquad
|h_o|=\frac{2|a|}{\big||b|-\sqrt D\big|}.
\]
Since $\big||b|-\sqrt D\big|\le |b|+\sqrt D$, $|h_o|\ge|h_2^*|$.
Finally,
\[
qh_2^*=\frac{D-b^2}{\operatorname{sgn}(b)(|b|+\sqrt D)}
=\operatorname{sgn}(b)(\sqrt D-|b|),
\]
so $b+qh_2^*=\operatorname{sgn}(b)\sqrt D$, proving (6). \hfill$\square$

\subsection*{Proof A.2: Theorem 2}
Substituting $\sqrt D=|b|\sqrt{1-u}$ into (5) yields
\[
h_2^*=-\frac{2a}{b\{1+\sqrt{1-u}\}}
=h_1^*\frac{2}{1+\sqrt{1-u}}.
\]
The multiplier is strictly positive for $u<1$. Taking logarithms of absolute values gives (7). \hfill$\square$

\subsection*{Proof A.3: Corollary 1}
Equation (7) writes $\Theta_{COT}$ exactly as the scalar composition $\phi\{u(a,b,q)\}$. Therefore any two triples $(a,b,q)$ having the same $u$ have the same COT value, and all derivatives or tests for $\Theta_{COT}$ can be expressed through $u$. \hfill$\square$

\subsection*{Proof A.4: Corollary 2}
Under $x\mapsto\alpha+\beta x$, $\beta\neq0$, the local derivatives transform as $(a,b,q)\mapsto(a,\beta b,\beta^2q)$, hence
$u\mapsto2(\beta^2q)a/(\beta^2b^2)=u$. Under response rescaling
$m-c\mapsto\lambda(m-c)$, $\lambda\neq0$, $(a,b,q)\mapsto\lambda(a,b,q)$ and again $u\mapsto u$. The result for $\Theta_{COT}=\phi(u)$ follows. \hfill$\square$

\subsection*{Proof A.5: Theorem 3}
Let $s=\sqrt{1-u}>0$. From (7),
$\phi(u)=\log2-\log(1+s)$ and $ds/du=-1/(2s)$, so
\[
\phi'(u)=\frac{1}{2s(1+s)}>0.
\]
At $u=0$, $s=1$ and $\phi(0)=0$. As $u\to-\infty$, $s\to\infty$ and $\phi(u)\to-\infty$; as $u\uparrow1$, $s\downarrow0$ and $\phi(u)\uparrow\log2$. Strict monotonicity and continuity give the stated bijection. Solving
$e^\Theta=2/(1+s)$ yields $s=2e^{-\Theta}-1$ and hence
$u=1-(2e^{-\Theta}-1)^2$. Allowing $s=0$ gives the continuous boundary value $\bar\phi(1)=\log2$. \hfill$\square$

\subsection*{Proof A.6: Corollary 3}
By Theorem 2, $|h_2^*/h_1^*|=e^{\phi(u)}$. Theorem 3 gives
$e^{\phi(u)}\in(0,2)$ for $u<1$. The upper endpoint is approached as
$u\uparrow1$ but is not attained in the regular domain; the lower endpoint is approached as $u\to-\infty$. Thus the first-order distance can understate the second-order distance by a factor approaching two, whereas the reciprocal ratio $|h_1^*/h_2^*|$ is unbounded. \hfill$\square$

\subsection*{Proof A.7: Corollary 4}
Theorem 3 shows that $\phi$ is strictly increasing and $\phi(0)=0$, hence
$\operatorname{sgn}\{\phi(u)\}=\operatorname{sgn}(u)$. Since
$u=2qa/b^2$ and $b^2>0$, $\operatorname{sgn}(u)=\operatorname{sgn}(qa)$. \hfill$\square$

\subsection*{Proof A.8: Proposition 1}
By Theorem 3, $\phi$ is injective and $\phi(0)=0$, so
$\Theta_{COT}=0$ if and only if $u=0$. With $a\neq0$ and $b\neq0$,
$u=2qa/b^2=0$ if and only if $q=0$. \hfill$\square$

\subsection*{Proof A.9: Theorem 4}
Use the integral remainder in (3). If $|m^{(3)}|\le M_3$ on the relevant interval, then
\[
|R(h)|\le \frac{M_3|h|^3}{6},\qquad
R'(h)=\int_0^h(h-t)m^{(3)}(x_0+t;\psi)\,dt,
\]
and therefore $|R'(h)|\le M_3h^2/2$. On
$J=[h_2^*-\Delta,h_2^*+\Delta]$,
\[
F_2'(h)=F_2'(h_2^*)+q(h-h_2^*).
\]
By (6), $|F_2'(h_2^*)|=\sqrt D$, so
$|F_2'(h)|\ge\sqrt D-|q|\Delta$. Also
$|h|\le|h_2^*|+\Delta$, hence
$|R'(h)|\le M_3(|h_2^*|+\Delta)^2/2$. The reverse triangle inequality gives
$|F'(h)|\ge L(\Delta)>0$ on $J$. Since $F'$ is continuous, it has constant sign on $J$, and $F$ is strictly monotone.
At the center, $F_2(h_2^*)=0$, so
$|F(h_2^*)|=|R(h_2^*)|\le M_3|h_2^*|^3/6\le L(\Delta)\Delta$.
If $F$ is increasing, the mean-value theorem gives
$F(h_2^*+\Delta)\ge F(h_2^*)+L(\Delta)\Delta\ge0$ and
$F(h_2^*-\Delta)\le F(h_2^*)-L(\Delta)\Delta\le0$; if $F$ is decreasing the inequalities reverse. In either case the intermediate value theorem gives a zero in $J$, and strict monotonicity makes it unique. \hfill$\square$

\subsection*{Proof A.10: Corollary 5}
When $|q|\Delta$ and $M_3(|h_2^*|+\Delta)^2/2$ are of smaller order than $\sqrt D$, $L(\Delta)=\sqrt D\{1+o(1)\}$. Theorem 4 therefore gives the leading-order displacement bound
\[
|h^*-h_2^*|\lesssim \frac{M_3|h_2^*|^3}{6\sqrt D}.
\]
Using $\sqrt D=|b|\sqrt{1-u}$ gives (10). Let the right-hand side equal
$\rho|h_2^*|$. If $\rho<1$, then $h^*$ and $h_2^*$ have the same sign and
$|h^*/h_2^*-1|\le\rho$. Consequently
$1-\rho\le|h^*/h_2^*|\le1+\rho$, so
\[
\left|\log\left|\frac{h^*}{h_1^*}\right|-\Theta_{COT}\right|
=\left|\log\left|\frac{h^*}{h_2^*}\right|\right|
\le\max\{-\log(1-\rho),\log(1+\rho)\}
=-\log(1-\rho).
\]
This is the stated leading-order log-scale bound. \hfill$\square$

\subsection*{Proof A.11: Theorem 5}
On $\{b\neq0,u<1\}$, $u(\eta)=2qa/b^2$ is continuously differentiable with gradient
\[
\nabla u(\eta)=\left(\frac{2q}{b^2},-\frac{4qa}{b^3},\frac{2a}{b^2}\right)^\top,
\]
and $\phi$ is continuously differentiable by Theorem 3. The multivariate delta method (van der Vaart, 1998) applied to $\phi\{u(\widehat\eta)\}$ yields
\[
\sqrt n(\widehat\Theta-\Theta_0)\Rightarrow
N\!\left(0,\phi'(u_0)^2\nabla u(\eta_0)^\top\Sigma\nabla u(\eta_0)\right).
\]
Consistency of the plug-in variance follows from $\widehat\eta\to_p\eta_0$, $\widehat\Sigma\to_p\Sigma$, and continuity. \hfill$\square$

\subsection*{Proof A.12: Theorem 6}
The delta method for $u(\widehat\eta_n)$ under the triangular array gives
$\sqrt n(\widehat u-u_n)\Rightarrow\omega Z$, $Z\sim N(0,1)$. Hence
\[
\sqrt n(1-\widehat u)=\sqrt n(1-u_n)-\sqrt n(\widehat u-u_n)
\Rightarrow V:=\kappa-\omega Z.
\]
Because $V$ has a continuous distribution,
$\Pr(\widehat u<1)\to\Pr(V>0)=\Phi(\kappa/\omega)$, proving (i).
Conditioning on the event $\widehat u<1$ therefore yields
$\sqrt n(1-\widehat u)\Rightarrow V\mid(V>0)$. By the continuous mapping theorem,
$n^{1/4}\sqrt{1-\widehat u}\Rightarrow\sqrt V$. From (7), with $s\downarrow0$,
$\log2-\phi(1-s^2)=\log(1+s)=s+O(s^2)$, so
$n^{1/4}(\log2-\widehat\Theta)\Rightarrow\sqrt V$, proving (ii).
Similarly,
$\phi'(u)=\{2\sqrt{1-u}(1+\sqrt{1-u})\}^{-1}$ implies
\[
n^{-1/4}\phi'(\widehat u)\Rightarrow\frac{1}{2\sqrt V}.
\]
Since $\widehat\omega\to_p\omega$,
$n^{1/4}\widehat{se}(\widehat\Theta)\Rightarrow\omega/(2\sqrt V)$.
Moreover $n^{1/4}(\log2-\Theta_n)\to\sqrt\kappa$, so
$n^{1/4}(\widehat\Theta-\Theta_n)\Rightarrow\sqrt\kappa-\sqrt V$.
Taking the ratio gives the non-Gaussian limit in part (iii). At $\kappa=0$, the conditioning event is $Z<0$, so $V=\omega|Z|$ with $|Z|$ half-normal. \hfill$\square$

\subsection*{Proof A.13: Theorem 7}
As $b\to0$, $D=b^2-2qa\to-2qa$. If $qa>0$, this limit is negative, giving no real quadratic root for sufficiently small $|b|$. If $qa<0$, then $u=2qa/b^2\to-\infty$ and
$\sqrt{1-u}=\sqrt{2|qa|}\,|b|^{-1}\{1+o(1)\}$. Theorem 2 therefore gives
\[
\left|\frac{h_2^*}{h_1^*}\right|
=\frac{2}{1+\sqrt{1-u}}
=\frac{2|b|}{\sqrt{2|qa|}}\{1+o(1)\}.
\]
Taking logarithms gives (12). Formula (5) converges to a finite nonzero value because its denominator converges in magnitude to $\sqrt{2|qa|}$. \hfill$\square$

\subsection*{Proof A.14: Proposition 2}
Theorem 3 gives the regular range $\Theta_{COT}\in(-\infty,\log2)$ and continuous closure $(-\infty,\log2]$. Therefore, if $\delta\ge\log2$, no regular principal-branch value can satisfy $\Theta_{COT}>\delta$, and the upper inadequacy set $\{\Theta_{COT}\ge\delta\}$ is empty in the regular parameter space. \hfill$\square$

\clearpage
\section*{References}\label{references}
\setlength{\parindent}{-0.5in}
\setlength{\leftskip}{0.5in}

Andrews, D. W. K. (1999). Estimation when a parameter is on a boundary. \emph{Econometrica, 67}(6), 1341--1383. \url{https://doi.org/10.1111/1468-0262.00082}

Andrews, D. W. K. (2000). Inconsistency of the bootstrap when a parameter is on the boundary of the parameter space. \emph{Econometrica, 68}(2), 399--405. \url{https://doi.org/10.1111/1468-0262.00114}

Bates, D. M., \& Watts, D. G. (1980). Relative curvature measures of nonlinearity. \emph{Journal of the Royal Statistical Society: Series B (Methodological), 42}(1), 1--16. \url{https://doi.org/10.1111/j.2517-6161.1980.tb01094.x}

Bates, D. M., \& Watts, D. G. (1988). \emph{Nonlinear regression analysis and its applications}. Wiley. \url{https://doi.org/10.1002/9780470316757}

Beale, E. M. L. (1960). Confidence regions in non-linear estimation. \emph{Journal of the Royal Statistical Society: Series B (Methodological), 22}(1), 41--76. \url{https://doi.org/10.1111/j.2517-6161.1960.tb00353.x}

Berger, R. L., \& Hsu, J. C. (1996). Bioequivalence trials, intersection-union tests and equivalence confidence sets. \emph{Statistical Science, 11}(4), 283--319. \url{https://doi.org/10.1214/ss/1032280304}

Box, M. J. (1971). Bias in nonlinear estimation. \emph{Journal of the Royal Statistical Society: Series B (Methodological), 33}(2), 171--190. \url{https://doi.org/10.1111/j.2517-6161.1971.tb00871.x}

Budtz-J{\o}rgensen, E., Keiding, N., \& Grandjean, P. (2001). Benchmark dose calculation from epidemiological data. \emph{Biometrics, 57}(3), 698--706. \url{https://doi.org/10.1111/j.0006-341x.2001.00698.x}

Chernoff, H. (1954). On the distribution of the likelihood ratio. \emph{The Annals of Mathematical Statistics, 25}(3), 573--578. \url{https://doi.org/10.1214/aoms/1177728725}

Crump, K. S. (1984). A new method for determining allowable daily intakes. \emph{Fundamental and Applied Toxicology, 4}(5), 854--871. \url{https://doi.org/10.1016/0272-0590(84)90107-6}

Davison, A. C., \& Hinkley, D. V. (1997). \emph{Bootstrap methods and their application}. Cambridge University Press. \url{https://doi.org/10.1017/CBO9780511802843}

Demidenko, E., Williams, B. B., Flood, A. B., \& Swartz, H. M. (2013). Standard error of inverse prediction for dose-response relationship: Approximate and exact statistical inference. \emph{Statistics in Medicine, 32}(12), 2048--2061. \url{https://doi.org/10.1002/sim.5668}

DiCiccio, T. J., \& Efron, B. (1996). Bootstrap confidence intervals. \emph{Statistical Science, 11}(3), 189--228. \url{https://doi.org/10.1214/ss/1032280214}

Efron, B., \& Tibshirani, R. J. (1993). \emph{An introduction to the bootstrap}. Chapman \& Hall/CRC. \url{https://doi.org/10.1201/9780429246593}

Fieller, E. C. (1954). Some problems in interval estimation. \emph{Journal of the Royal Statistical Society: Series B (Methodological), 16}(2), 175--185. \url{https://doi.org/10.1111/j.2517-6161.1954.tb00159.x}

Gleser, L. J., \& Hwang, J. T. (1987). The nonexistence of $100(1-\alpha)\%$ confidence sets of finite expected diameter in errors-in-variables and related models. \emph{The Annals of Statistics, 15}(4), 1351--1362. \url{https://doi.org/10.1214/aos/1176350597}

Gregory, A. W., \& Veall, M. R. (1985). Formulating Wald tests of nonlinear restrictions. \emph{Econometrica, 53}(6), 1465--1468. \url{https://doi.org/10.2307/1913221}

Hall, P. (1992). \emph{The bootstrap and Edgeworth expansion}. Springer. \url{https://doi.org/10.1007/978-1-4612-4384-7}

Hirschberg, J., \& Lye, J. (2010). A geometric comparison of the delta and Fieller confidence intervals. \emph{The American Statistician, 64}(3), 234--241. \url{https://doi.org/10.1198/tast.2010.08130}

Lafontaine, F., \& White, K. J. (1986). Obtaining any Wald statistic you want. \emph{Economics Letters, 21}(1), 35--40. \url{https://doi.org/10.1016/0165-1765(86)90117-5}

Osborne, C. (1991). Statistical calibration: A review. \emph{International Statistical Review, 59}(3), 309--336. \url{https://doi.org/10.2307/1403690}

Phillips, P. C. B., \& Park, J. Y. (1988). On the formulation of Wald tests of nonlinear restrictions. \emph{Econometrica, 56}(5), 1065--1083. \url{https://doi.org/10.2307/1911359}

Ratkowsky, D. A. (1983). \emph{Nonlinear regression modeling: A unified practical approach}. Marcel Dekker.

Ritz, C., Baty, F., Streibig, J. C., \& Gerhard, D. (2015). Dose-response analysis using R. \emph{PLOS ONE, 10}(12), e0146021. \url{https://doi.org/10.1371/journal.pone.0146021}

Jensen, S. M., Kluxen, F. M., \& Ritz, C. (2019). A review of recent advances in benchmark dose methodology. \emph{Risk Analysis, 39}(10), 2295--2315. \url{https://doi.org/10.1111/risa.13324}

Lin, L., Piegorsch, W. W., \& Bhattacharya, R. (2015). Nonparametric benchmark dose estimation with continuous dose-response data. \emph{Scandinavian Journal of Statistics, 42}(3), 713--731. \url{https://doi.org/10.1111/sjos.12132}

Piegorsch, W. W. (2014). Model uncertainty in environmental dose-response risk analysis. \emph{Statistics and Public Policy, 1}(1), 78--85. \url{https://doi.org/10.1080/2330443X.2014.937021}

Ritz, C., \& Streibig, J. C. (2005). Bioassay analysis using R. \emph{Journal of Statistical Software, 12}(5), 1--22. \url{https://doi.org/10.18637/jss.v012.i05}

Seber, G. A. F., \& Wild, C. J. (1989). \emph{Nonlinear regression}. Wiley. \url{https://doi.org/10.1002/0471725315}

Schuirmann, D. J. (1987). A comparison of the two one-sided tests procedure and the power approach for assessing the equivalence of average bioavailability. \emph{Journal of Pharmacokinetics and Biopharmaceutics, 15}(6), 657--680. \url{https://doi.org/10.1007/BF01068419}

Self, S. G., \& Liang, K.-Y. (1987). Asymptotic properties of maximum likelihood estimators and likelihood ratio tests under nonstandard conditions. \emph{Journal of the American Statistical Association, 82}(398), 605--610. \url{https://doi.org/10.1080/01621459.1987.10478472}

Shao, K., \& Gift, J. S. (2014). Model uncertainty and Bayesian model averaged benchmark dose estimation for continuous data. \emph{Risk Analysis, 34}(1), 101--120. \url{https://doi.org/10.1111/risa.12078}

van der Vaart, A. W. (1998). \emph{Asymptotic statistics}. Cambridge University Press. \url{https://doi.org/10.1017/CBO9780511802256}

Wellek, S. (2010). \emph{Testing statistical hypotheses of equivalence and noninferiority} (2nd ed.). Chapman \& Hall/CRC. \url{https://doi.org/10.1201/EBK1439808184}

Wheeler, M. W., \& Bailer, A. J. (2007). Properties of model-averaged BMDLs: A study of model averaging in dichotomous response risk estimation. \emph{Risk Analysis, 27}(3), 659--670. \url{https://doi.org/10.1111/j.1539-6924.2007.00920.x}

Wheeler, M. W., \& Bailer, A. J. (2008). Model averaging software for dichotomous dose response risk estimation. \emph{Journal of Statistical Software, 26}(5), 1--15. \url{https://doi.org/10.18637/jss.v026.i05}

Wheeler, M. W., Cortinas, J., Aerts, M., Gift, J. S., \& Davis, J. A. (2022). Continuous model averaging for benchmark dose analysis: Averaging over distributional forms. \emph{Environmetrics, 33}(5), e2728. \url{https://doi.org/10.1002/env.2728}

\end{document}